\documentclass[preprint2]{aastex}

\newdimen\minuswidth    
\setbox0=\hbox{$-$}
\minuswidth=\wd0
\catcode`@=\active
\def@{\kern\minuswidth}
\newdimen\digitwidth    
\setbox0=\hbox{\rm0}
\digitwidth=\wd0
\catcode`!=\active
\def!{\kern\digitwidth}

\shorttitle{Pre-Main Sequence Stars  found in N66/NGC346}
\shortauthors{A. Nota et al.}

\begin{document}


\title{The Discovery of a Population of Pre-Main Sequence Stars in N66/NGC346 from Deep
HST/ACS Images\footnote{Based on observations with the NASA/ESA Hubble Space Telescope,
obtained at the Space Telescope Science Institute, which is operated by AURA Inc., under NASA
contract NAS 5-26555}}


\author{A. Nota\footnote{Space Telescope Science Institute, 3700 San Martin Drive, Baltimore,
MD 21218, USA,nota@stsci.edu} \altaffilmark{,3},
M. Sirianni\altaffilmark{2,}\footnote{ESA, Space Telescope Operations Division},
E. Sabbi\altaffilmark{2},
M. Tosi\altaffilmark{}\footnote{INAF--Osservatorio Astronomico di Bologna, via Ranzani 1,
Bologna, I--40127,  Italy},
M. Clampin\altaffilmark{}\footnote{Goddard Space Flight Center, Greenbelt, Maryland, MD
20771, USA},
J. Gallagher\altaffilmark{}\footnote{Department of Astronomy, University of Wisconsin, 475
North Charter Street, Madison, WI 53706-1582, USA},
M. Meixner\altaffilmark{2},
S. Oey\altaffilmark{}\footnote{Department of Astronomy, University of Michigan, 830 Dennison
Buidling, Ann Arbor, Michigan, MI 48109, USA},
A. Pasquali\altaffilmark{}\footnote{Max-Planck-Institut fuer Astronomie, Koenigstuhl 17,
D-69117 Heidelberg, Germany },
L. J. Smith\altaffilmark{}\footnote{Department of Physics and Astronomy, University College
London, Gower Street, London, WC1E 6BT, GB},
R. Walterbos\altaffilmark{}\footnote{Department of Astronomy, New Mexico State University,
Department 4500, P.O. Box 30001, Las Cruces, NM 88003, USA}, 
J. Mack\altaffilmark{2}}



\begin{abstract}
We report the discovery of a rich population of low mass stars in the young,
massive star forming region N66/NGC346 in the Small Magellanic Cloud, from deep
V, I and H$\alpha$ images taken with the HST/ACS. These stars have likely formed
together with the NGC346 cluster, $\simeq$ 3-5 Myr ago. 
Their magnitude and colors are those of pre-main sequence stars in the mass range
0.6-3 M$_\odot$, mostly concentrated in the main cluster, but with secondary
subclusters spread over a region across $\sim 45$ pc. These subclusters appear to
be spatially coincident with previously known knots of molecular gas identified
in ground based and ISO observations. We show that N66/NGC346 is a complex
region, being shaped by its massive stars, and the observations presented here
represent a key step towards the understanding of how star formation occurred and
has progressed in this low metallicity environment.
\end{abstract}





\section{Introduction}

N66 is the most massive and active star forming region in the Small Magellanic
Cloud  (SMC) and  the star cluster NGC346 is its ionizing core. It contains more
than half of the known O stars in this galaxy, and associated CO clouds
\citep[][]{rubio00}. The upper end of its stellar mass distribution has been well
studied in the recent years: very early O--type stars (O3) and high mass (45--100
M$_\odot$) stars were found suggesting a slope of the Initial Mass Function (IMF)
in NGC346 very similar to the slope for massive stars in LMC and solar
neighborhood ($\Gamma$ = {\it d}(log $\xi$) /{\it d}(log M) = -1.9), and an age
of $\sim 3$ Myr \citep[see][and reference there in]{bouret03}.
\citet{MPG} also suggested  presence of sequential star formation in NGC346, with
star formation beginning at the southwest side of the association, and spreading
towards the central cluster.

While a wealth of information is available on the upper end of the stellar mass
distribution in N66/NGC346, little is known of the low end of the distribution.
The SMC is characterized by a dust-to-gass mass ratio that is 30 times lower than
in the Galaxy \citep{stan00}, and a metallicity about 1/5 than that of the
Galaxy. May these differences have had an impact on the cluster formation
mechanism and their distribution of stellar masses? To answer this question we
began a comprehensive investigation of star formation in the SMC, from the ground
and with HST. This paper presents the first results on N66/NGC346.

\section{Observations}

We used the HST/ACS Wide Field Channel (WF) to map NGC346 in the filters F555W
($\sim$V), F814W ($\sim$I), and F658N ($\sim$ H$\alpha$).  The observations were
carried out between July 13 and 17, 2004, as part of Cycle 13 proposal 10248 
(PI. A. Nota).  In the broad band filters, nine deep images (200'' $\times$ 200''
each) were obtained in each filter, partially overlapping to get optimal spatial
coverage of the NGC346 region. An additional set of short exposures was taken
in the same instrumental configuration to perform accurate photometry of the
brightest stars.  A single image of NGC346 was taken in the light of H$\alpha$,
with an exposure time of 1542s sec.

In Plate~1, we show the fully reduced, multidrizzled and combined V--I color
composite image of NGC 346. The image covers an area of $\simeq$ 5' $\times$ 5'
which corresponds to $\simeq 87 \times 87$ pc, at  a distance to the SMC of 59.7
Kpc \citep{dolphin01}. The exposure time in the central region is $\simeq$ 4100
sec in each filter. In Plate~2 we show the  fully reduced H$\alpha$ image.

All images were reduced using the standard STScI ACS pipeline CALACS (HST Data
Handbook for ACS).

Photometry was performed on the drizzled images by aperture and PSF fitting,
using the DAOPHOT package within IRAF. Stars were automatically detected with the
routine DAOFIND in all the frames, with a detection threshold set at 4$\sigma$
above the local background level. Features misinterpreted for stars, such as
noise spikes, were rejected. A spatially variable PSF was computed for both V and
I images, using$\sim 140$ isolated stars in different positions of the images. No
CTE correction was applied. However, preliminary CTE determinations indicate that
such correction is small, and can therefore be neglected in this first paper
without changing the overall conclusions. Our final photometry is calibrated in
the Vegamag system, with the zeropoints adopted from Sirianni et al. (2005) of
25.724 and 25.501 for the V and I filters respectively. 

\section{The morphology of the NGC346 region}

The deep HST/ACS images show  an unprecedented level of detail in the NGC346/N66
complex. The NGC346 cluster at the center is resolved into at least three
subclusters, and a myriad of small compact clusters is visible throughout the
region.  Some of these clusters appear to be still embedded in dust and
nebulosity, and they might be sites of recent or ongoing star formation. The
diffuse nebulosity is due to the presence of H$\beta$ and  [OIII] 4959, 5007 in
the V filter, which indicate the bright underlying HII region. This extends to
the N with arches and ragged filaments, but is delineated to the S-SW by a sharp
fan-like {\it ridge}, which encompasses the entire cluster.  This ridge is
especially evident in the H$\alpha$ image of the region (Plate~2). Extensive dust
lanes are well visible along this {\it ridge}. Dust patches are also visible
throughout the region, most likely the remnant of the original molecular cloud
\citep[][]{mizuno01}. 


\section{The Color-Magnitude Diagram}

\begin{figure}
\includegraphics[angle=0,scale=5]{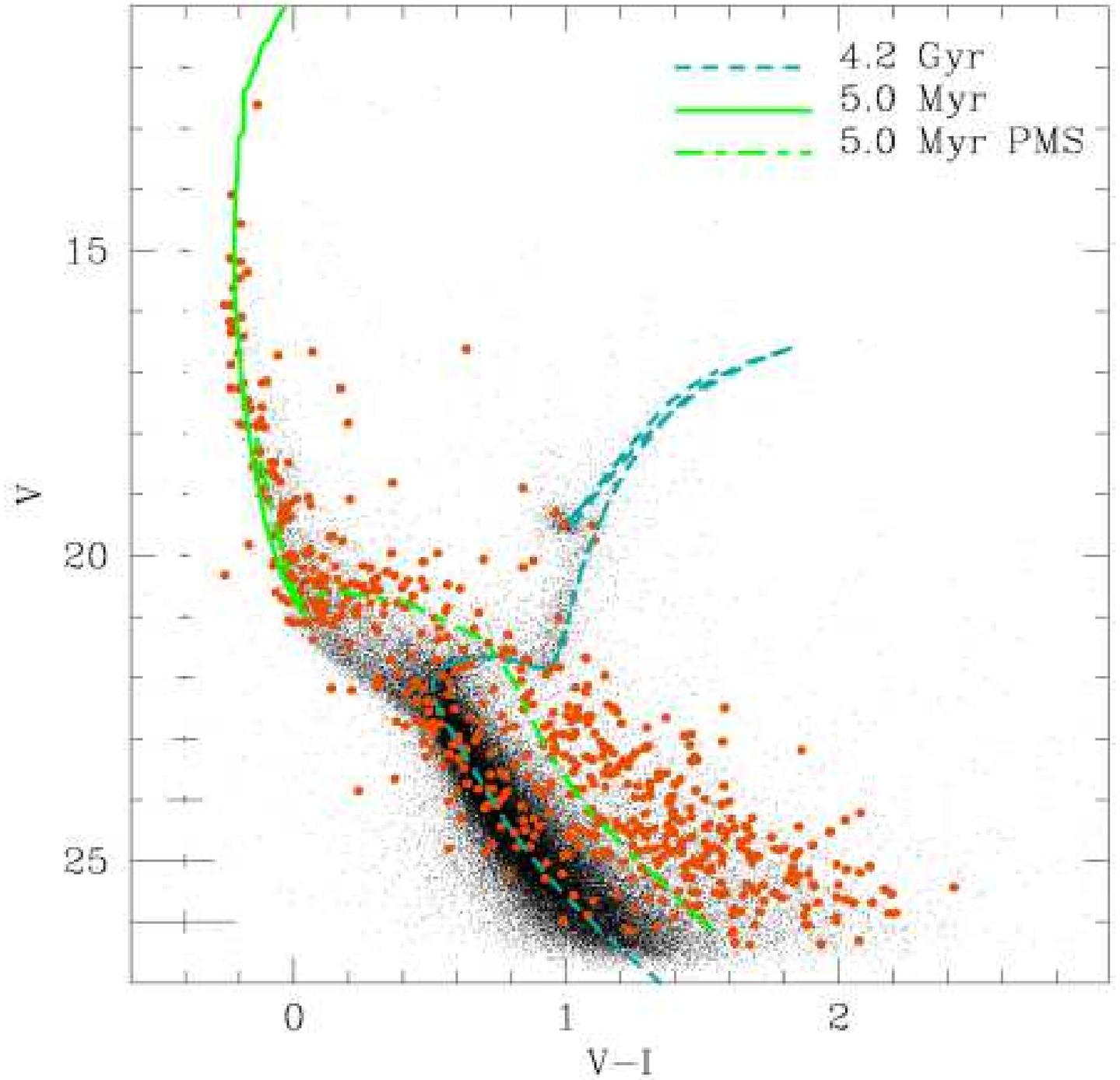}
\caption{The Color Magnitude Diagram of the NGC346  central cluster (large orange dots)
superimposed to the CMD of the entire region (small dots). Two isochrones  (5 Myr and 4.2
Gyr) are  overplotted for reference.\label{fig4}}
\end{figure}

The V and I photometric data with photometric error smaller than 0.1 mag was used
to construct the V, V-I Color-Magnitude Diagram (CMD) of the entire NGC346
region. The CMD (small black dots in Fig.~\ref{fig4}) contains $\simeq$ 80000
stars. Thanks to the high photometric accuracy obtained with the ACS, these stars
produce a very clean CMD, with a well defined narrow main sequence (MS), down to
V $\simeq$ 26.  The photometric errors in magnitude and color are indicated
directly in the CMD, as side bars. A pronounced old main sequence turnoff is
clearly visible, at V $\simeq$ 22, and so is the red clump, at V $\simeq$ 19.5.
The CMD morphology also reveals a young population, indicated by the bright, blue
MS, extending from the MS of the old SMC stellar population.  Below V $\simeq$ 21
mag a group of very red faint stars is detected to the right of the MS. 
Plausibly, this split MS could be due to differential reddening. However, the
absence of such a split in the upper part of the MS led us to believe that these
are stars that are less massive than $\simeq$3 M$_{\odot}$ (down to 0.6
M$_{\odot}$), which formed with the rest of the  central cluster but have not
reached the MS yet (pre-MS stars).

Analyzing separately the stellar content of the individual clusters we resolve in
the ACS images, we can better characterize the local stellar population. The
orange dots in Figure~\ref{fig4} highlight the observational CMD for the core of
NGC346. They indicate stars within a radius of $\simeq$ 8$''$ from the cluster
center. This is superimposed to the CMD of the entire region (small black dots). 

We superimposed to these CMDs isochrones for pre-MS stars from
\citet[][]{siess00}, and for MS and post-MS stars from \citet[][]{lejeune01}, and
\citet[][]{pietrinferni04}, to define the age  and the nature of the population
observed in the central cluster. We find presence of at least two different
populations: 
\begin{enumerate}
\item a composite SMC  field population covering a wide range in age, but
dominated by the 4  Gyr component,
\item a very young population of age $\simeq$3-5 Myr. Stars less massive than
$\simeq$3 M$_{\odot}$ (down to 0.6 M$_{\odot}$) are still in the pre-MS phase,
approaching the MS.
\end{enumerate}

An intermediate age population ($\simeq$150 Myr) might also be present. 

\section{The {\it candidate} pre-MS stars}

\begin{figure}
\includegraphics[angle=0,scale=5]{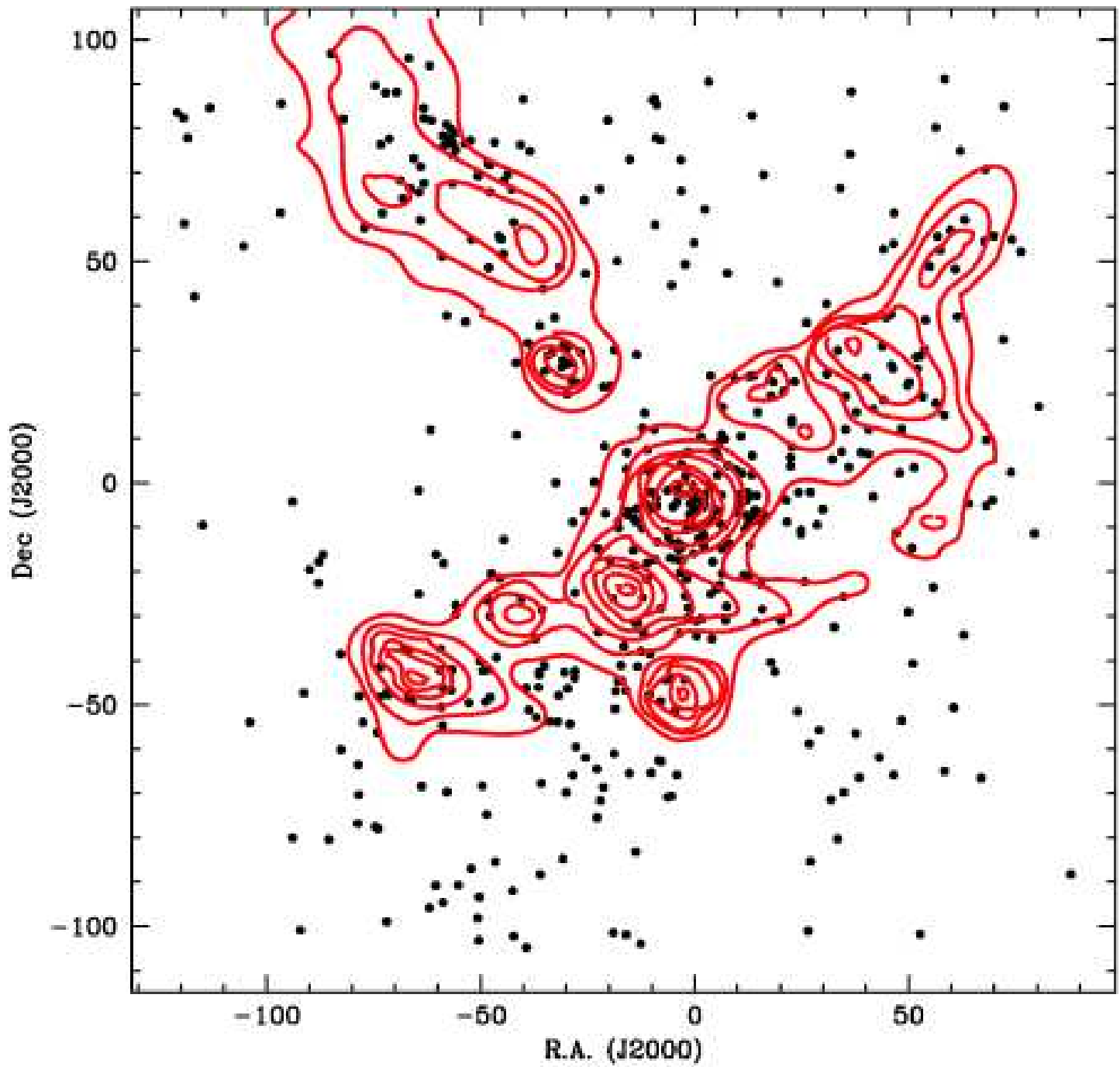}
\caption{The spatial distribution of the pre-MS stars (black dots) shown superimposed to the
7$\mu m$ emission peaks from ISO/LW2 \citep[from Fig. 8 of][]{rubio00}\label{fig5}}
\end{figure}

The pre-MS stars are, as expected, faint (V $<$ 21) and  red (1.5$< $V-I $<$2.2).
The CMD isochrone fit indicates they likely formed with the rest of the cluster
($\simeq$ 3-5 Myr ago), but have not reached the MS yet. They are in the mass
range 0.6 - 3 M$_\odot$. Their spatial distribution is shown in
Figure~\ref{fig5}. They appear concentrated at the center of NGC346, mostly in
the central subclusters,  but also in the small compact clusters that are located
on a direction which is perpendicular to the NGC346 main body, extending to the
NE. Their spatial location appears coincident with the clumps of neutral material
identified by
\citet{rubio00}.
The spatial density of the pre-MS stars appear to decrease as we move from the
center towards the S-SW {\it ridge}, where we observe the dust lanes. As also
suggested by \citet[][]{rubio00}, it is likely that the region, and especially
the {\it ridge}, may be  a site of  more recent or on going star formation, and 
the forming stars are still embedded in dust and too faint to be visually
detected in the ACS images.  

\section{Star formation in NGC346}

The region around N66/NGC346 is a remarkable region that has displayed a high
level of activity in the recent past. To the NE of N66/NGC346 there is a known SN
remnant, SNR0057-7226, first identified as a bright source in {\it Einstein}
X-ray observations \citep[][]{inoue83}. 
From the expansion velocity derived from FUSE observations of $\simeq$150 km
s$^{-1}$ \citep[][]{danforth03}, we can  derive a dynamical timescale of 2
$\times$ 10$^{5}$ yr and set a lower limit to the epoch of the SN explosion.  It
is clear that SNR0057-7226 has  been interacting with the surrounding region, and
the O VI CIII and X-ray emission observed \citep{danforth03} arises from the
shock interaction.  
Two concentric bubbles are seen to the NE in the H$\alpha$ region shown in
Plate~2: we believe they are most likely shaped by the powerful winds of HD5980.

Observations of \citet[][]{rubio00} and \citet[][]{contursi00} have painted a
detailed picture of the molecular  gas and dust component in  N66/NGG346. 
Their observations indicate that the molecular gas has mostly been dissociated by
the UV radiation of the hottest stars: what has not been dissociated is detected
in small, compact clumps, which appear to be spatially concident with the regions
where we observe the pre-main sequence stars  \citep[][]{rubio00}. Specifically,
their CO(2-1) SEST observations indicate  clumps of emission in the NGC346
central cluster, and, even stronger emission along the perpendicular direction to
the NGC346 main body, to the NE. These clumps are exactly where we find the
highest concentration of pre-MS stars. There appear not to be much molecular gas
or dust outside the NGC346 main body, casting the doubt that the  SW {\it ridge}
might not be  at the boundary with a molecular cloud as initially suspected
\citep[][]{rubio00}.   

\bigskip

We have shown that N66/NGC346  is a complex region that has been shaped by the
evolution of the most massive stars, some of which have already exploded as
supernova and are compressing the central region from different directions. The
outstanding question will be to understand how star formation has been triggered
and has been progressing in this region. Further necessary work on this region
includes spectroscopic characterization of the pre-main sequence stars, and
spectroscopic observations of the gas kinematics. On the theoretical side, NGC346
can easily become the ideal laboratory to test current theories for  star
formation at low masses.


\acknowledgments

Financial support was provided by the Italian MIUR to ES and MT, through Cofin 2002028935 and
2004020323, and STScI GO Grant GO-10248.07-A.




\begin{small}

\end{small}
\clearpage



\clearpage

\clearpage

\begin{figure}
\includegraphics[angle=0,scale=5]{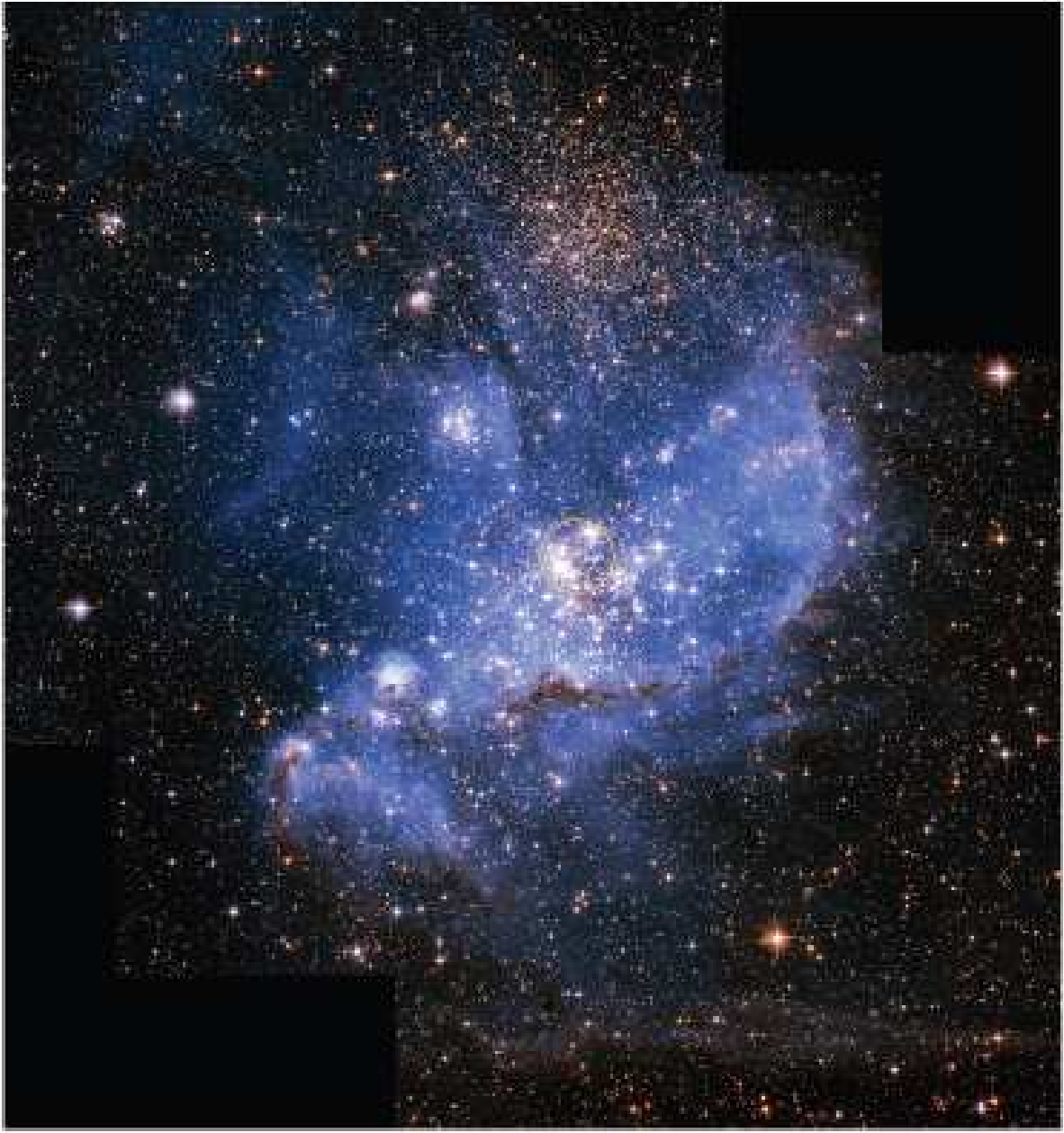}
\caption{{\it Plate--1} VI color composite of three partially overlapping HST/ACS WF images of
NGC346, covering a total area of 5' $\times$ 5'. North is up and East to the
left. The yellow circle (radius$=$8\arcsec) marks the position on NGC346 \label{fig1}}
\end{figure}

\clearpage

\begin{figure}
\includegraphics[angle=0,scale=8]{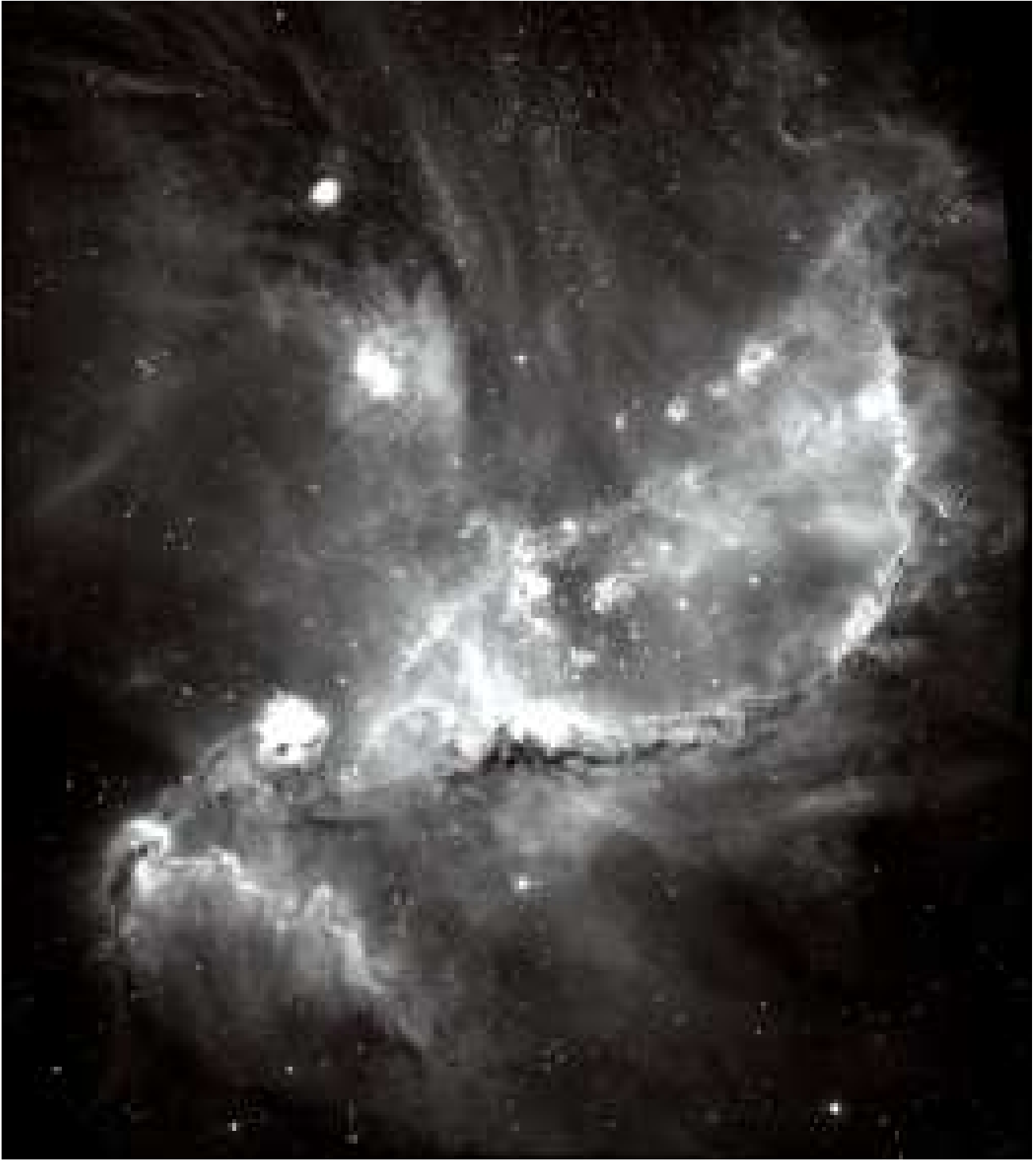}
\caption{{\it Plate--2} H$\alpha$ HST/ACS WF image of  the central NGC346 region, covering an
area of 200$''$ $\times$ 200 $''$. North is up and East to the left.\label{fig2}}
\end{figure}


\end{document}